# RELATIVELY INERTIAL DELAYS[1]

Serban E. Vlad

Oradea City Hall

**Abstract**

*The paper studies the relatively inertial delays that represent one of the most important concepts in the modeling of the asynchronous circuits.*

## 1. Introduction

The delays are the mathematical models of the delay circuits. Delay theory is the mathematical theory that considers the fundamental circuit in digital electronics be the delay circuit and modeling is made by using delays and Boolean functions. The most detailed level of modeling is considered, starting from the delays that occur in gates and wires.

Relative inertia is the property of the states of having their speed of variation limited by the persistency of the input and the relatively inertial delays are these delays the states of which are relatively inertial. Even if the concept has close connections with the published literature, it has a severe shortcoming: the serial connection of two relatively inertial delays is not always a relatively inertial delay.

Some major properties of these delays are presented as well as the relation with absolute inertia and with zenoness.

## 2. Preliminaries

**Definition 2.1** The binary Boole algebra is the set $\boldsymbol{B} = \{0,1\}$ endowed with the discrete topology, with the order $0 \leq 1$ and with the usual laws $\overline{\phantom{x}}, \cdot, \cup, \oplus$.

**Definition 2.2** The order and the laws of $\boldsymbol{B}$ induce an order and laws having the same notations in the set of the $\boldsymbol{R} \to \boldsymbol{B}$ functions.

**Definition 2.3** Let $x : \boldsymbol{R} \to \boldsymbol{B}$ and $A \subset \boldsymbol{R}$ be given. We define
$$\bigcap_{\xi \in A} x(\xi) = \begin{cases} 0, if\ \exists \xi \in A, x(\xi) = 0 \\ 1, otherwise \end{cases}$$

---

**Definition 2.4** The left limit function $x(t-0)$ of $x: \mathbf{R} \to \mathbf{B}$ is defined by
$$\forall t \in \mathbf{R}, \exists \varepsilon > 0, \forall \xi \in (t-\varepsilon, t), x(\xi) = x(t-0)$$

**Definition 2.5** The functions $\overline{x(t-0)} \cdot x(t), x(t-0) \cdot \overline{x(t)}$ are called the left semi-derivatives of $x$.

**Definition 2.6** For the function $x: \mathbf{R} \to \mathbf{B}$, we define $\lim_{t \to \infty} x(t) \in \mathbf{B}$ by
$$\exists t_f \in \mathbf{R}, \forall \xi \geq t_f, x(\xi) = \lim_{t \to \infty} x(t)$$

**Definition 2.7** We use the notation $\chi_A : \mathbf{R} \to \mathbf{B}$ for the characteristic function of the set $A \subset \mathbf{R}$:
$$\chi_A(t) = \begin{cases} 1, & \text{if } t \in A \\ 0, & \text{otherwise} \end{cases}$$

A function $x: \mathbf{R} \to \mathbf{B}$ for which an unbounded sequence $t_0 < t_1 < t_2 < ...$ of real numbers exists so that
$$x(t) = x(t_0) \cdot \chi_{(-\infty, t_1)}(t) \oplus x(t_1) \cdot \chi_{[t_1, t_2)}(t) \oplus x(t_2) \cdot \chi_{[t_2, t_3)}(t) \oplus ...$$
is called signal. The set of the signals is denoted by $S$.

**Definition 2.8** A multi-valued function $f: S \to P^*(S)$, where $P^*(S) = \{X \mid X \subset S, X \neq \emptyset\}$ is called system (with 1-dimensional inputs and 1-dimensional states). Any $u \in S$ is called input and any $x \in f(u)$ is called state.

**Definition 2.9** A delay is a system $f$ satisfying the next property
$$\forall u \in S, \exists \lim_{t \to \infty} u(t) \Rightarrow \forall x \in f(u), \exists \lim_{t \to \infty} x(t) \text{ and } \lim_{t \to \infty} x(t) = \lim_{t \to \infty} u(t)$$

## 3. Relative inertia

**Definition 3.1** We consider the next inequalities:
$$\begin{cases} \overline{x(t-0)} \cdot x(t) \leq \bigcap_{\xi \in [t-\delta_r, t-\delta_r + \mu_r]} u(\xi) \\ x(t-0) \cdot \overline{x(t)} \leq \bigcap_{\xi \in [t-\delta_f, t-\delta_f + \mu_f]} \overline{u(\xi)} \end{cases} \quad (1)$$

where $0 \leq \mu_r \leq \delta_r$, $0 \leq \mu_f \leq \delta_f$ and $u, x \in S$. The system $f_{RI}^{\mu_r, \delta_r, \mu_f, \delta_f} : S \to P^*(S)$ that is defined by

$$\forall u \in S, f_{RI}^{\mu_r,\delta_r,\mu_f,\delta_f}(u) = \{x \mid x \in S, x \text{ fulfills } (1)\}$$

is called the relative inertia property.

**Definition 3.2** If the systems $f$ and $f_{RI}^{\mu_r,\delta_r,\mu_f,\delta_f}$ satisfy

$$\forall u \in S, f(u) \subset f_{RI}^{\mu_r,\delta_r,\mu_f,\delta_f}(u)$$

then $f$ is called relatively inertial. We use to say that $f$ satisfies the relative inertia property $f_{RI}^{\mu_r,\delta_r,\mu_f,\delta_f}$.

## 4. Order

**Definition 4.1** The order $\subset$ of the systems $f, g : S \to P^*(S)$ is
$$f \subset g \Leftrightarrow \forall u \in S, f(u) \subset g(u)$$

**Theorem 4.2** For any numbers $0 \le \mu_r \le \delta_r$, $0 \le \mu_f \le \delta_f$, $0 \le \mu'_r \le \delta'_r$, $0 \le \mu'_f \le \delta'_f$:

$$f_{RI}^{\mu_r,\delta_r,\mu_f,\delta_f} \subset f_{RI}^{\mu'_r,\delta'_r,\mu'_f,\delta'_f} \Leftrightarrow$$
$$\delta_r \ge \delta'_r, \delta_f \ge \delta'_f, \delta'_r - \mu'_r \ge \delta_r - \mu_r, \delta'_f - \mu'_f \ge \delta_f - \mu_f$$

**Proof** $f_{RI}^{\mu_r,\delta_r,\mu_f,\delta_f} \subset f_{RI}^{\mu'_r,\delta'_r,\mu'_f,\delta'_f} \Leftrightarrow$

$$\Leftrightarrow \forall t \in \mathbf{R}, \forall u \in S, \bigcap_{\xi \in [t-\delta_r, t-\delta_r+\mu_r]} u(\xi) \le \bigcap_{\xi \in [t-\delta'_r, t-\delta'_r+\mu'_r]} u(\xi) \text{ and}$$
$$\text{and} \bigcap_{\xi \in [t-\delta_f, t-\delta_f+\mu_f]} u(\xi) \le \bigcap_{\xi \in [t-\delta'_f, t-\delta'_f+\mu'_f]} u(\xi)$$

$$\Leftrightarrow \forall t \in \mathbf{R}, [t-\delta_r, t-\delta_r+\mu_r] \supset [t-\delta'_r, t-\delta'_r+\mu'_r] \text{ and}$$
$$\text{and } [t-\delta_f, t-\delta_f+\mu_f] \supset [t-\delta'_f, t-\delta'_f+\mu'_f]$$

$$\Leftrightarrow \delta_r \ge \delta'_r, \delta_r - \mu_r \le \delta'_r - \mu'_r, \delta_f \ge \delta'_f, \delta_f - \mu_f \le \delta'_f - \mu'_f$$

**Theorem 4.3** If $f$ is a delay, then any system $g \subset f$ is a delay.

**Proof** Let $u \in S$ be arbitrary s.t. $\lim_{t \to \infty} u(t) = \lambda$. Because the hypothesis states

$$\forall x \in f(u), \exists \lim_{t \to \infty} x(t) \text{ and } \lim_{t \to \infty} x(t) = \lambda$$

$$g(u) \subset f(u)$$

we have
$$\forall x \in g(u), \exists \lim_{t \to \infty} x(t) \text{ and } \lim_{t \to \infty} x(t) = \lambda$$

**Remark 4.4** When $f \subset f_{RI}^{\mu_r, \delta_r, \mu_f, \delta_f}$ is a relatively inertial delay, any system $g \subset f$ is a relatively inertial delay $\subset f_{RI}^{\mu_r, \delta_r, \mu_f, \delta_f}$.

**Definition 4.5** The relative inertia property $f_{RI}^{\mu_r, \delta_r, \mu_f, \delta_f}$ generated by the delay $f$ is defined by: $f \subset f_{RI}^{\mu_r, \delta_r, \mu_f, \delta_f}$ and for any $f_{RI}^{\mu_r', \delta_r', \mu_f', \delta_f'}$ with $f \subset f_{RI}^{\mu_r', \delta_r', \mu_f', \delta_f'}$ we have $f_{RI}^{\mu_r, \delta_r, \mu_f, \delta_f} \subset f_{RI}^{\mu_r', \delta_r', \mu_f', \delta_f'}$.

## 5. Duality

**Definition 5.1** The dual system $f^*$ of the system $f$ is defined like this
$$\forall u \in S, f^*(u) = \{\overline{x} \mid x \in f(\overline{u})\}$$

**Theorem 5.2** The dual of $f_{RI}^{\mu_r, \delta_r, \mu_f, \delta_f}$ is $f_{RI}^{\mu_f, \delta_f, \mu_r, \delta_r}$.

**Proof** Let $u \in S$ an arbitrary input for which we can write
$$(f_{RI}^{\mu_r, \delta_r, \mu_f, \delta_f})^*(u) = \{\overline{x} \mid x \in f_{RI}^{\mu_r, \delta_r, \mu_f, \delta_f}(\overline{u})\}$$
$$= \{\overline{x} \mid \overline{x(t-0)} \cdot x(t) \leq \bigcap_{\xi \in [t-\delta_r, t-\delta_r+\mu_r]} \overline{u}(\xi), x(t-0) \cdot \overline{x(t)} \leq \bigcap_{\xi \in [t-\delta_f, t-\delta_f+\mu_f]} u(\xi) \}$$
$$= \{x \mid x(t-0) \cdot \overline{x(t)} \leq \bigcap_{\xi \in [t-\delta_r, t-\delta_r+\mu_r]} \overline{u}(\xi), \overline{x(t-0)} \cdot x(t) \leq \bigcap_{\xi \in [t-\delta_f, t-\delta_f+\mu_f]} u(\xi) \}$$
$$= f_{RI}^{\mu_f, \delta_f, \mu_r, \delta_r}(u)$$

**Theorem 5.3** The system $f$ is a delay iff the dual system $f^*$ is a delay.
**Proof**
$$\forall u \in S, \exists \lim_{t \to \infty} u(t) \Rightarrow \forall x \in f(u), \exists \lim_{t \to \infty} x(t) \text{ and } \lim_{t \to \infty} x(t) = \lim_{t \to \infty} u(t)$$
$$\Leftrightarrow (\forall \overline{u} \in S, \exists \lim_{t \to \infty} \overline{u}(t) \Rightarrow \forall \overline{x} \in f(\overline{u}), \exists \lim_{t \to \infty} \overline{x}(t) \text{ and } \lim_{t \to \infty} \overline{x}(t) = \lim_{t \to \infty} \overline{u}(t))$$

$$\Leftrightarrow (\forall u \in S, \exists \lim_{t\to\infty} u(t) \Rightarrow \forall x \in f^*(u), \exists \lim_{t\to\infty} x(t) \text{ and } \lim_{t\to\infty} x(t) = \lim_{t\to\infty} u(t))$$

where we have used the notations $\forall \bar{u} \in S \equiv \forall u, \bar{u} \in S$ and $\forall \bar{x} \in f(\bar{u}) \equiv \forall x, \bar{x} \in f(\bar{u})$.

**Theorem 5.4** Let $f$ be a delay. Then
$$f \subset f_{RI}^{\mu_r, \delta_r, \mu_f, \delta_f} \Leftrightarrow f^* \subset f_{RI}^{\mu_f, \delta_f, \mu_r, \delta_r}$$

**Proof** $f \subset f_{RI}^{\mu_r, \delta_r, \mu_f, \delta_f} \Leftrightarrow \forall u \in S, f(u) \subset f_{RI}^{\mu_r, \delta_r, \mu_f, \delta_f}(u)$

$\Leftrightarrow \forall u \in S, \{x \mid x \in f(u)\} \subset \{x \mid x \in f_{RI}^{\mu_r, \delta_r, \mu_f, \delta_f}(u)\}$

$\Leftrightarrow \forall \bar{u} \in S, \{\bar{x} \mid x \in f(\bar{u})\} \subset \{\bar{x} \mid x \in f_{RI}^{\mu_r, \delta_r, \mu_f, \delta_f}(\bar{u})\}$

$\Leftrightarrow \forall u \in S, f^*(u) \subset (f_{RI}^{\mu_r, \delta_r, \mu_f, \delta_f})^*(u)$

$\Leftrightarrow \forall u \in S, f^*(u) \subset f_{RI}^{\mu_f, \delta_f, \mu_r, \delta_r}(u) \Leftrightarrow f^* \subset f_{RI}^{\mu_f, \delta_f, \mu_r, \delta_r}$

We have use the notation $\forall \bar{u} \in S \equiv \forall u, \bar{u} \in S$ and the result from Theorem 5.2.

## 6. Serial connection

**Definition 6.1** The serial connection $g \circ f$ of the systems $f, g$ is defined like this:
$$\forall u \in S, (g \circ f)(u) = \bigcup_{x \in f(u)} g(x)$$

**Remark 6.2** The serial connection of the relative inertia properties is not a relative inertia property in general and the serial connection of the relatively inertial delays is not necessarily a relatively inertial delay. We give the example of the next delays

$$\begin{cases} x(t-0) \cdot x(t) = \bigcap_{\xi \in [t-2,t)} \overline{x(\xi)} \cdot u(t) \\ x(t-0) \cdot \overline{x(t)} = \bigcap_{\xi \in [t-2,t)} x(\xi) \cdot \overline{u(t)} \end{cases}, \begin{cases} \overline{y(t-0)} \cdot y(t) = \bigcap_{\xi \in [t-4,t)} \overline{y(\xi)} \cdot x(t) \\ y(t-0) \cdot \overline{y(t)} = \bigcap_{\xi \in [t-4,t)} y(\xi) \cdot \overline{x(t)} \end{cases}$$

that are relatively inertial, but their serial connection is not a relatively inertial delay. The fact that the first system is a delay is shown by supposing that $\lim_{t\to\infty} u(t) = \lambda, \lambda \in B$ thus $t_f \in R$ exists s.t. $\forall t \geq t_f, u(t) = \lambda$ and consequently

$$\forall t \geq t_f, \begin{cases} \overline{x(t-0)} \cdot x(t) = \bigcap_{\xi \in [t-2,t)} \overline{x(\xi)} \cdot \lambda \\ x(t-0) \cdot \overline{x(t)} = \bigcap_{\xi \in [t-2,t)} x(\xi) \cdot \overline{\lambda} \end{cases}$$

We take $\lambda = 1$, in other words $\forall t \geq t_f, x(t)$ can switch from 0 to 1 exactly once. If $x(t_f) = 1$, then $\forall t \geq t_f, x(t) = 1$ and if $x(t_f) = 0$, then

$$\exists t_1 \in (t_f, t_f + 2], \overline{x(t_1-0)} \cdot x(t_1) = \bigcap_{\xi \in [t_1-2,t_1)} \overline{x(\xi)} \cdot 1 = 1$$

$$\forall t \geq t_1, x(t) = 1$$

The presumption that $\lambda = 0$ brings the conclusion that $\lim_{t \to \infty} x(t) = 0$.

The proof that

$$\begin{cases} \overline{y(t-0)} \cdot y(t) = \bigcap_{\xi \in [t-4,t)} \overline{y(\xi)} \cdot x(t) \\ y(t-0) \cdot \overline{y(t)} = \bigcap_{\xi \in [t-4,t)} y(\xi) \cdot \overline{x(t)} \end{cases}$$

is a delay is similar.

The two delays are relatively inertial because

$$\begin{cases} \overline{x(t-0)} \cdot x(t) \leq u(t) \\ x(t-0) \cdot \overline{x(t)} \leq \overline{u(t)} \end{cases}, \begin{cases} \overline{y(t-0)} \cdot y(t) \leq x(t) \\ y(t-0) \cdot \overline{y(t)} \leq \overline{x(t)} \end{cases}$$

For reasons of symmetry, by supposing against all reason that their serial connection would be relatively inertial, it would satisfy

$$\begin{cases} \overline{y(t-0)} \cdot y(t) \leq \bigcap_{\xi \in [t-\delta, t-\delta+\mu]} u(\xi) \\ y(t-0) \cdot \overline{y(t)} \leq \bigcap_{\xi \in [t-\delta, t-\delta+\mu]} \overline{u(\xi)} \end{cases} \quad (2)$$

where $0 \leq \mu \leq \delta$. We choose $u(t) = \chi_{[0,1) \cup [2,3) \cup [4,\infty)}(t)$ for which the first delay gives $x(t) = \chi_{[0,3) \cup [5,\infty)}(t)$ and the second delay shows us that $y(t) = \chi_{[0,4) \cup [8,\infty)}(t)$. We have

$$\bigcap_{\xi \in [t-\delta, t-\delta+\mu]} u(\xi) = \chi_{[\delta, 1+\delta-\mu) \cup [2+\delta, 3+\delta-\mu) \cup [4+\delta, \infty)}(t)$$

$$\bigcap_{\xi \in [t-\delta, t-\delta+\mu]} \overline{u(\xi)} = \chi_{(-\infty, \delta-\mu) \cup [1+\delta, 2+\delta-\mu) \cup [3+\delta, 4+\delta-\mu)}(t)$$

where the intervals $[\delta, 1+\delta-\mu)$, $[2+\delta, 3+\delta-\mu)$, $[1+\delta, 2+\delta-\mu)$, $[3+\delta, 4+\delta-\mu)$ can be empty or non-empty in principle. From (2) we obtain

$$\chi_{\{0,8\}}(t) \leq \chi_{[\delta,1+\delta-\mu)\cup[2+\delta,3+\delta-\mu)\cup[4+\delta,\infty)}(t) \qquad (3)$$

$$\chi_{\{4\}}(t) \leq \chi_{(-\infty,\delta-\mu)\cup[1+\delta,2+\delta-\mu)\cup[3+\delta,4+\delta-\mu)}(t) \qquad (4)$$

(3) implies $\delta \leq 0$, thus $\delta = \mu = 0$. This represents a contradiction with (4) that becomes

$$\chi_{\{4\}}(t) \leq \chi_{(-\infty,0)\cup[1,2)\cup[3,4)}(t)$$

impossible. (2) is false.

**Theorem 6.3** The serial connection of the delays is a delay.
**Proof** The hypothesis states that

$$\forall u \in S, \exists \lim_{t\to\infty} u(t) \Rightarrow \forall x \in f(u), \exists \lim_{t\to\infty} x(t) \text{ and } \lim_{t\to\infty} x(t) = \lim_{t\to\infty} u(t) \qquad (5)$$

$$\forall x \in S, \exists \lim_{t\to\infty} x(t) \Rightarrow \forall y \in g(x), \exists \lim_{t\to\infty} y(t) \text{ and } \lim_{t\to\infty} y(t) = \lim_{t\to\infty} x(t) \qquad (6)$$

and let $u \in S$ be an arbitrary input s.t. $\lim_{t\to\infty} u(t) = \lambda, \lambda \in B$. From (5) we infer

$$\forall x \in f(u), \exists \lim_{t\to\infty} x(t) \text{ and } \lim_{t\to\infty} x(t) = \lambda \qquad (7)$$

We fix an arbitrary $x \in f(u)$. From (6) and (7) we draw the conclusion that

$$\forall y \in g(x), \exists \lim_{t\to\infty} y(t) \text{ and } \lim_{t\to\infty} y(t) = \lambda$$

We have just proved that

$$\forall u \in S, \exists \lim_{t\to\infty} u(t) \Rightarrow \forall y \in (g \circ f)(u), \exists \lim_{t\to\infty} y(t) \text{ and } \lim_{t\to\infty} y(t) = \lim_{t\to\infty} u(t)$$

i.e. $g \circ f$ is a delay.

## 7. Intersection

**Definition 7.1** Let $f, g$ be two systems. If $\forall u \in S, f(u) \cap g(u) \neq \emptyset$, then the intersection $f \cap g$ of $f$ and $g$ is by definition the next system:

$$\forall u \in S, (f \cap g)(u) = f(u) \cap g(u)$$

**Theorem 7.2** For any $0 \leq \mu_r \leq \delta_r$, $0 \leq \mu_f \leq \delta_f$, $0 \leq \mu'_r \leq \delta'_r$, $0 \leq \mu'_f \leq \delta'_f$ the system $f_{RI}^{\mu_r,\delta_r,\mu_f,\delta_f} \cap f_{RI}^{\mu'_r,\delta'_r,\mu'_f,\delta'_f}$ is relatively inertial $\subset f_{RI}^{\mu_r,\delta_r,\mu_f,\delta_f}$.

**Proof** We observe that $f_{RI}^{\mu_r,\delta_r,\mu_f,\delta_f} \cap f_{RI}^{\mu'_r,\delta'_r,\mu'_f,\delta'_f}$ is described by the equations

$$\begin{cases} \overline{x(t-0)} \cdot x(t) \leq \bigcap_{\xi \in [t-\delta_r, t-\delta_r+\mu_r]} u(\xi) \cdot \bigcap_{\xi \in [t-\delta_r', t-\delta_r'+\mu_r']} u(\xi) \\ x(t-0) \cdot \overline{x(t)} \leq \bigcap_{\xi \in [t-\delta_f, t-\delta_f+\mu_f]} \overline{u(\xi)} \cdot \bigcap_{\xi \in [t-\delta_f', t-\delta_f'+\mu_f']} \overline{u(\xi)} \end{cases}$$

that have always a solution, for example the constant function $x(t) = \lambda$. In this moment the theorem is proved, but let us remark also for all $t \in R$ and all $u \in S$ the truth of the inequalities

$$\bigcap_{\xi \in [t-\delta_r, t-\delta_r+\mu_r]} u(\xi) \cdot \bigcap_{\xi \in [t-\delta_r', t-\delta_r'+\mu_r']} u(\xi) \leq \bigcap_{\xi \in [t-\delta_r, t-\delta_r+\mu_r]} u(\xi)$$

$$\bigcap_{\xi \in [t-\delta_f, t-\delta_f+\mu_f]} \overline{u(\xi)} \cdot \bigcap_{\xi \in [t-\delta_f', t-\delta_f'+\mu_f']} \overline{u(\xi)} \leq \bigcap_{\xi \in [t-\delta_f, t-\delta_f+\mu_f]} \overline{u(\xi)}$$

**Theorem 7.3** If the delays $f, g$ satisfy $\forall u \in S, f(u) \cap g(u) \neq \emptyset$, then $f \cap g$ is a delay.

**Proof** The fact that $f \cap g \subset f$ and that $f$ is a delay shows from Theorem 4.3 that $f \cap g$ is a delay.

**Theorem 7.4** We suppose that the relatively inertial delays $f \subset f_{RI}^{\mu_r, \delta_r, \mu_f, \delta_f}$, $g \subset f_{RI}^{\mu_r', \delta_r', \mu_f', \delta_f'}$ satisfy $\forall u \in S, f(u) \cap g(u) \neq \emptyset$. Then $f \cap g \subset f_{RI}^{\mu_r, \delta_r, \mu_f, \delta_f}$ is a relatively inertial delay.

**Proof** $f \cap g$ is a delay from Theorem 7.3. We have

$$f \cap g \subset f_{RI}^{\mu_r, \delta_r, \mu_f, \delta_f} \cap f_{RI}^{\mu_r', \delta_r', \mu_f', \delta_f'} \subset f_{RI}^{\mu_r, \delta_r, \mu_f, \delta_f}$$

**Theorem 7.5** If the delay $f$ satisfies $\forall u \in S, f(u) \cap f_{RI}^{\mu_r, \delta_r, \mu_f, \delta_f}(u) \neq \emptyset$, then $f \cap f_{RI}^{\mu_r, \delta_r, \mu_f, \delta_f} \subset f_{RI}^{\mu_r, \delta_r, \mu_f, \delta_f}$ is a relatively inertial delay.

**Proof** $f \cap f_{RI}^{\mu_r, \delta_r, \mu_f, \delta_f}$ is a delay because it is a subsystem of $f$ and we take into account Theorem 4.3. The statement is obvious.

## 8. Union

**Definition 8.1** The union $f \cup g$ of the systems $f, g$ is defined in the next way:

$$\forall u \in S, (f \cup g)(u) = f(u) \cup g(u)$$

**Remark 8.2** The union of the relative inertia properties is not a relative inertia property in general and similarly the union of the relatively inertial delays is not a relatively inertial delay in general. We give the example of the union of the next relative inertia properties

$$\begin{cases} \overline{x(t-0)} \cdot x(t) \leq \bigcap_{\xi \in [t-3,t-2]} u(\xi) \\ x(t-0) \cdot \overline{x(t)} \leq \bigcap_{\xi \in [t-3,t-2]} \overline{u(\xi)} \end{cases}, \begin{cases} \overline{x(t-0)} \cdot x(t) \leq \bigcap_{\xi \in [t-1,t]} u(\xi) \\ x(t-0) \cdot \overline{x(t)} \leq \bigcap_{\xi \in [t-1,t]} \overline{u(\xi)} \end{cases}$$

that is given by

$$\begin{cases} \overline{x(t-0)} \cdot x(t) \leq \bigcap_{\xi \in [t-3,t-2]} u(\xi) \cup \bigcap_{\xi \in [t-1,t]} u(\xi) \\ x(t-0) \cdot \overline{x(t)} \leq \bigcap_{\xi \in [t-3,t-2]} \overline{u(\xi)} \cup \bigcap_{\xi \in [t-1,t]} \overline{u(\xi)} \end{cases}$$

and it is not a relative inertia property. By presuming against all reason that it is a relative inertia property, for reasons of symmetry we must have

$$\begin{cases} \bigcap_{\xi \in [t-3,t-2]} u(\xi) \cup \bigcap_{\xi \in [t-1,t]} u(\xi) \leq \bigcap_{\xi \in [t-\delta,t-\delta+\mu]} u(\xi) \\ \bigcap_{\xi \in [t-3,t-2]} \overline{u(\xi)} \cup \bigcap_{\xi \in [t-1,t]} \overline{u(\xi)} \leq \bigcap_{\xi \in [t-\delta,t-\delta+\mu]} \overline{u(\xi)} \end{cases} \quad (8)$$

where $0 \leq \mu \leq \delta$. For $u = \chi_{[0,2)}$ we infer that

$$\chi_{[1,2) \cup [3,4)}(t) = \bigcap_{\xi \in [t-3,t-2]} u(\xi) \cup \bigcap_{\xi \in [t-1,t]} u(\xi) \leq \bigcap_{\xi \in [t-\delta,t-\delta+\mu]} u(\xi) = \chi_{[\delta, 2+\delta-\mu)}(t) \quad (9)$$

$$\chi_{(-\infty,2) \cup [3,\infty)}(t) = \bigcap_{\xi \in [t-3,t-2]} \overline{u(\xi)} \cup \bigcap_{\xi \in [t-1,t]} \overline{u(\xi)} \leq \quad (10)$$

$$\leq \bigcap_{\xi \in [t-\delta,t-\delta+\mu]} \overline{u(\xi)} = \chi_{(-\infty,\delta-\mu) \cup [2+\delta,\infty)}(t)$$

The conclusion is the next one:

$$2 \stackrel{(10)}{\leq} \delta - \mu \leq \delta \stackrel{(9)}{\leq} 1$$

contradiction from which we get that the inequalities (8) are false.

**Theorem 8.3** The union of the delays is a delay.
**Proof** The hypothesis states that

$$\forall u \in S, \exists \lim_{t \to \infty} u(t) \Rightarrow \forall x \in f(u), \exists \lim_{t \to \infty} x(t) \text{ and } \lim_{t \to \infty} x(t) = \lim_{t \to \infty} u(t)$$

$$\forall u \in S, \exists \lim_{t \to \infty} u(t) \Rightarrow \forall x \in g(u), \exists \lim_{t \to \infty} x(t) \text{ and } \lim_{t \to \infty} x(t) = \lim_{t \to \infty} u(t)$$

and let $u \in S$ be arbitrary with $\lim_{t \to \infty} u(t) = \lambda$. We infer that

$$\forall x \in f(u) \cup g(u), \exists \lim_{t \to \infty} x(t) \text{ and } \lim_{t \to \infty} x(t) = \lambda$$

thus $f \cup g$ is a delay.

## 9. Non-anticipation

**Definition 9.1** We denote by $u_{|A}$ the restriction of $u \in S$ to $A \subset \mathbf{R}$. The system $f$ is called non-anticipatory if:

$$\forall t \in \mathbf{R}, \forall u \in S, \forall v \in S,$$
$$u_{|(-\infty,t]} = v_{|(-\infty,t]} \Rightarrow \{x_{|(-\infty,t]} \mid x \in f(u)\} = \{y_{|(-\infty,t]} \mid y \in f(v)\}$$

**Theorem 9.2** For any parameters $0 \le \mu_r \le \delta_r$, $0 \le \mu_f \le \delta_f$ the relative inertia property $f_{RI}^{\mu_r,\delta_r,\mu_f,\delta_f}$ is non-anticipatory.

**Proof** The property is true because in the inequalities

$$\begin{cases} \overline{x(t-0)} \cdot x(t) \le \bigcap_{\xi \in [t-\delta_r, t-\delta_r+\mu_r]} u(\xi) \\ x(t-0) \cdot \overline{x(t)} \le \bigcap_{\xi \in [t-\delta_f, t-\delta_f+\mu_f]} \overline{u(\xi)} \end{cases}$$

we have $t - \delta_r + \mu_r \le t, t - \delta_f + \mu_f$ thus the solutions $x \in f_{RI}^{\mu_r,\delta_r,\mu_f,\delta_f}(u)$ depend only on the values $u_{(-\infty,t]}$.

**Theorem 9.3** Let $f$ be a non-anticipatory delay and we suppose that $\forall u \in S, f(u) \cap f_{RI}^{\mu_r,\delta_r,\mu_f,\delta_f}(u) \ne \emptyset$. Then the relatively inertial delay $f \cap f_{RI}^{\mu_r,\delta_r,\mu_f,\delta_f}$ is non-anticipatory.

**Proof** Let $t \in \mathbf{R}$, $u, v \in S$ be arbitrary s.t. $u_{|(-\infty,t]} = v_{|(-\infty,t]}$. From the fact that $f \cap f_{RI}^{\mu_r,\delta_r,\mu_f,\delta_f}$ exists and from

$$\{x_{|(-\infty,t]} \mid x \in f(u)\} = \{y_{|(-\infty,t]} \mid y \in f(v)\}$$
$$\{x_{|(-\infty,t]} \mid x \in f_{RI}^{\mu_r,\delta_r,\mu_f,\delta_f}(u)\} = \{y_{|(-\infty,t]} \mid y \in f_{RI}^{\mu_r,\delta_r,\mu_f,\delta_f}(v)\}$$

we infer

$$\{x_{|(-\infty,t]} \mid x \in (f \cap f_{RI}^{\mu_r,\delta_r,\mu_f,\delta_f})(u)\} =$$

$$= \{x_{|(-\infty,t]} \mid x \in f(u) \cap f_{RI}^{\mu_r,\delta_r,\mu_f,\delta_f}(u)\} =$$

$$= \{x_{|(-\infty,t]} \mid x \in f(u)\} \cap \{x_{|(-\infty,t]} \mid x \in f_{RI}^{\mu_r,\delta_r,\mu_f,\delta_f}(u)\} =$$

$$= \{y_{|(-\infty,t]} \mid y \in f(v)\} \cap \{y_{|(-\infty,t]} \mid y \in f_{RI}^{\mu_r,\delta_r,\mu_f,\delta_f}(v)\} =$$

$$= \{y_{|(-\infty,t]} \mid y \in f(v) \cap f_{RI}^{\mu_r,\delta_r,\mu_f,\delta_f}(v)\} =$$

$$= \{y_{|(-\infty,t]} \mid y \in (f \cap f_{RI}^{\mu_r,\delta_r,\mu_f,\delta_f})(v)\}$$

## 10. Time invariance

**Definition 10.1** We denote by $\tau^d : R \to R$ the translation with $d \in R$: $\forall t \in R, \tau^d(t) = t - d$. The system $f$ is time invariant if

$$\forall d \in R, \forall u \in S, f(u \circ \tau^d) = \{x \circ \tau^d \mid x \in f(u)\}$$

**Theorem 10.2** The relative inertia property $f_{RI}^{\mu_r,\delta_r,\mu_f,\delta_f}$ is time invariant.

**Proof** We fix $d \in R$ and $u \in S$ arbitrarily. We can write that

$$f_{RI}^{\mu_r,\delta_r,\mu_f,\delta_f}(u \circ \tau^d) = \left\{ x \mid \begin{array}{l} \overline{x(t-0)} \cdot x(t) \leq \bigcap\limits_{\xi \in [t-\delta_r, t-\delta_r+\mu_r]} u(\xi-d) \\ x(t-0) \cdot \overline{x(t)} \leq \bigcap\limits_{\xi \in [t-\delta_f, t-\delta_f+\mu_f]} \overline{u(\xi-d)} \end{array} \right\}$$

$$\stackrel{(\xi'=\xi-d)}{=} \left\{ x \mid \begin{array}{l} \overline{x(t-0)} \cdot x(t) \leq \bigcap\limits_{\xi'+d \in [t-\delta_r, t-\delta_r+\mu_r]} u(\xi') \\ x(t-0) \cdot \overline{x(t)} \leq \bigcap\limits_{\xi'+d \in [t-\delta_f, t-\delta_f+\mu_f]} \overline{u(\xi')} \end{array} \right\}$$

$$= \left\{ x \mid \begin{array}{l} \overline{x(t-0)} \cdot x(t) \leq \bigcap\limits_{\xi \in [t-d-\delta_r, t-d-\delta_r+\mu_r]} u(\xi) \\ x(t-0) \cdot \overline{x(t)} \leq \bigcap\limits_{\xi \in [t-d-\delta_f, t-d-\delta_f+\mu_f]} \overline{u(\xi)} \end{array} \right\}$$

$$= \left\{ x \mid \begin{array}{l} \overline{x(t+d-0)} \cdot x(t+d) \leq \bigcap\limits_{\xi \in [t-\delta_r, t-\delta_r+\mu_r]} u(\xi) \\ x(t+d-0) \cdot \overline{x(t+d)} \leq \bigcap\limits_{\xi \in [t-\delta_f, t-\delta_f+\mu_f]} \overline{u(\xi)} \end{array} \right\}$$

$$= \{x \mid x \circ \tau^{-d} \in f_{RI}^{\mu_r,\delta_f,\mu_f,\delta_f}(u)\} = \{x \circ \tau^d \mid x \in f_{RI}^{\mu_r,\delta_f,\mu_f,\delta_f}(u)\}$$

**Theorem 10.3** Let $f$ be a time invariant delay. If $\forall u \in S$, $f(u) \cap f_{RI}^{\mu_r,\delta_r,\mu_f,\delta_f}(u) \neq \emptyset$, then the relatively inertial delay $f \cap f_{RI}^{\mu_r,\delta_r,\mu_f,\delta_f}$ is time invariant.

**Proof** Let $d \in R$ and $u \in S$ be arbitrary. We have:

$$(f \cap f_{RI}^{\mu_r,\delta_f,\mu_f,\delta_f})(u \circ \tau^d) = f(u \circ \tau^d) \cap f_{RI}^{\mu_r,\delta_f,\mu_f,\delta_f}(u \circ \tau^d) =$$

$$= \{x \circ \tau^d \mid x \in f(u)\} \cap \{x \circ \tau^d \mid x \in f_{RI}^{\mu_r,\delta_f,\mu_f,\delta_f}(u)\} =$$

$$= \{x \circ \tau^d \mid x \in f(u) \cap f_{RI}^{\mu_r,\delta_f,\mu_f,\delta_f}(u)\} =$$

$$= \{x \circ \tau^d \mid x \in (f \cap f_{RI}^{\mu_r,\delta_f,\mu_f,\delta_f})(u)\}$$

## 11. Absolute inertia

**Definition 11.1** We consider the next inequalities where $d_r \geq 0, d_f \geq 0$ and $x \in S$:

$$\begin{cases} \overline{x(t-0)} \cdot x(t) \leq \bigcap_{\xi \in [t,t+d_r]} x(\xi) \\ x(t-0) \cdot \overline{x(t)} \leq \bigcap_{\xi \in [t,t+d_f]} \overline{x(\xi)} \end{cases} \quad (11)$$

The absolute inertia property is the system $f_{AI}^{d_r,d_f}$ that is defined by

$$\forall u \in S, f_{AI}^{d_r,d_f}(u) = \{x \mid x \in S, x \text{ satisfies } (11)\}$$

**Definition 11.2** If the system $f$ fulfills

$$\forall u \in S, f(u) \subset f_{AI}^{d_r,d_f}(u)$$

then it is called an absolutely inertial system and we use to say that $f$ satisfies the absolute inertia property $f_{AI}^{d_r,d_f}$.

**Theorem 11.3** Let $0 \leq \mu_r \leq \delta_r$, $0 \leq \mu_f \leq \delta_f$ be arbitrary. If $\delta_f \geq \delta_r - \mu_r$, $\delta_r \geq \delta_f - \mu_f$ then $\forall u \in S, f_{RI}^{\mu_r,\delta_r,\mu_f,\delta_f}(u) \subset f_{AI}^{\delta_f-\delta_r+\mu_r,\delta_r-\delta_f+\mu_f}(u)$.

**Proof** We presume that $\delta_f \geq \delta_r - \mu_r, \delta_r \geq \delta_f - \mu_f$ and let $u, x \in S$ have the property that

$$\begin{cases} \overline{x(t-0)} \cdot x(t) \leq \bigcap_{\xi \in [t-\delta_r, t-\delta_r+\mu_r]} u(\xi) \\ x(t-0) \cdot \overline{x(t)} \leq \bigcap_{\xi \in [t-\delta_f, t-\delta_f+\mu_f]} \overline{u(\xi)} \end{cases}$$

We take $t_1 < t_2$ two time instants where $x$ switches in the next manner:
$$\overline{x(t_1-0)} \cdot x(t_1) = 1 \text{ and } x(t_2-0) \cdot \overline{x(t_2)} = 1$$

We obtain

$$\bigcap_{\xi \in [t_1-\delta_r, t_1-\delta_r+\mu_r]} u(\xi) = \bigcap_{\xi \in [t_2-\delta_f, t_2-\delta_f+\mu_f]} \overline{u(\xi)} = 1$$

$$\Rightarrow [t_1 - \delta_r, t_1 - \delta_r + \mu_r] \cap [t_2 - \delta_f, t_2 - \delta_f + \mu_f] = \varnothing$$

$$\Rightarrow t_1 - \delta_r + \mu_r < t_2 - \delta_f \text{ or } t_2 - \delta_f + \mu_f < t_1 - \delta_r$$

$$\Rightarrow t_2 - t_1 > \delta_f - \delta_r + \mu_r \text{ or } t_2 - t_1 < \delta_f - \mu_f - \delta_r$$

$$\Rightarrow t_2 - t_1 > \delta_f - \delta_r + \mu_r$$

(the inequality $t_2 - t_1 < \delta_f - \mu_f - \delta_r$ is false because the left term is strictly positive and the right term is non-positive.) The conclusion is that
$$\overline{x(t_1-0)} \cdot x(t_1) \leq \bigcap_{\xi \in [t_1, t_1+\delta_f-\delta_r+\mu_r]} x(\xi)$$

The other inequality is similarly shown.

**Remark 11.4** The fact that the delay $f$ is relatively inertial $f \subset f_{RI}^{\mu_r, \delta_r, \mu_f, \delta_f}$ and that $\delta_f \geq \delta_r - \mu_r$, $\delta_r \geq \delta_f - \mu_f$ implies from Theorem 11.3 that $f$ is absolutely inertial $f \subset f_{AI}^{\delta_f - \delta_r + \mu_r, \delta_r - \delta_f + \mu_f}$.

## 12. Zeno delays

**Definition 12.1** If the system $f$ satisfies one of the properties
  i) $\forall \varepsilon > 0, \exists t \in \mathbf{R}, \exists t' \in \mathbf{R}, \exists u \in S, \exists x \in f(u),$
  $$\overline{x(t-0)} \cdot x(t) = 1 \text{ and } \overline{x(t'-0)} \cdot x(t') = 1 \text{ and } 0 < t'-t < \varepsilon$$
  ii) $\forall \varepsilon > 0, \exists t \in \mathbf{R}, \exists t' \in \mathbf{R}, \exists u \in S, \exists x \in f(u),$
  $$x(t-0) \cdot \overline{x(t)} = 1 \text{ and } x(t'-0) \cdot \overline{x(t')} = 1 \text{ and } 0 < t'-t < \varepsilon$$
then it is called Zeno.

**Theorem 12.2** The system $f$ is not Zeno iff $\exists d_r > 0, \exists d_f > 0, f \subset f_{AI}^{d_r, d_f}$.

**Proof** $f$ is not Zeno $\Leftrightarrow$ the next properties are satisfied

$\exists d_r > 0, \forall t \in \mathbf{R}, \forall t' \in \mathbf{R}, \forall u \in S, \forall x \in f(u),$

$\quad (t < t' \text{ and } \overline{x(t-0)} \cdot x(t) = 1 \text{ and } x(t'-0) \cdot \overline{x(t')} = 1) \Rightarrow t'-t \geq d_r$

$\exists d_f > 0, \forall t \in \mathbf{R}, \forall t' \in \mathbf{R}, \forall u \in S, \forall x \in f(u),$

$\quad (t < t' \text{ and } x(t-0) \cdot \overline{x(t)} = 1 \text{ and } \overline{x(t'-0)} \cdot x(t') = 1) \Rightarrow t'-t \geq d_f$

$\Leftrightarrow$ the next properties are satisfied

$\exists d_r > 0, \forall t \in \mathbf{R}, \forall t' \in \mathbf{R}, \forall u \in S, \forall x \in f(u),$

$\quad (t < t' \text{ and } \overline{x(t-0)} \cdot x(t) = 1 \text{ and } x(t'-0) \cdot \overline{x(t')} = 1) \Rightarrow t'-t > d_r$

$\exists d_f > 0, \forall t \in \mathbf{R}, \forall t' \in \mathbf{R}, \forall u \in S, \forall x \in f(u),$

$\quad (t < t' \text{ and } x(t-0) \cdot \overline{x(t)} = 1 \text{ and } \overline{x(t'-0)} \cdot x(t') = 1) \Rightarrow t'-t > d_f$

$\Leftrightarrow$ the next properties are satisfied

$\exists d_r > 0, \forall t \in \mathbf{R}, \forall u \in S, \forall x \in f(u),$

$$\overline{x(t-0)} \cdot x(t) \leq \bigcap_{\xi \in [t, t+d_r]} x(\xi)$$

$\exists d_f > 0, \forall t \in \mathbf{R}, \forall u \in S, \forall x \in f(u),$

$$x(t-0) \cdot \overline{x(t)} \leq \bigcap_{\xi \in [t, t+d_f]} \overline{x(\xi)}$$

$\Leftrightarrow$ the next properties are satisfied

$$\exists d_r > 0, \forall u \in S, f(u) \subset f_{AI}^{d_r, 0}(u)$$

$$\exists d_f > 0, \forall u \in S, f(u) \subset f_{AI}^{0, d_f}(u)$$

$\Leftrightarrow \exists d_r > 0, \exists d_f > 0, \forall u \in S, f(u) \subset (f_{AI}^{d_r, 0} \cap f_{AI}^{0, d_f})(u)$

$\Leftrightarrow \exists d_r > 0, \exists d_f > 0, f \subset f_{AI}^{d_r, d_f}$

**Corollary 12.3** $f_{RI}^{\mu_r, \delta_r, \mu_f, \delta_f}$ is not Zeno iff $\delta_f > \delta_r - \mu_r, \delta_r > \delta_f - \mu_f$.

**Proof** If $d_r > 0, d_f > 0$ exist, namely $d_r = \delta_f - \delta_r + \mu_r$, $d_f = \delta_r - \delta_f + \mu_f$ s.t. $f_{RI}^{\mu_r, \delta_r, \mu_f, \delta_f} \subset f_{AI}^{d_r, d_f}$ (see Theorem 11.3). We apply Theorem 12.2.

<u>Only if</u> The hypothesis states that $f_{RI}^{\mu_r,\delta_r,\mu_f,\delta_f}$ is not Zeno. We suppose against all reason that $\delta_f \leq \delta_r - \mu_r$ and we take $u = \chi_{(-\infty,0)}$, for which $x \in f_{RI}^{\mu_r,\delta_r,\mu_f,\delta_f}(u)$ is equivalent with

$$\begin{cases} \overline{x(t-0)} \cdot x(t) \leq \bigcap_{\xi \in [t-\delta_r, t-\delta_r+\mu_r]} u(\xi) = \chi_{(-\infty, \delta_r - \mu_r)}(t) \\ x(t-0) \cdot \overline{x(t)} \leq \bigcap_{\xi \in [t-\delta_f, t-\delta_f+\mu_f]} \overline{u(\xi)} = \chi_{[\delta_f, \infty)}(t) \end{cases}$$

We observe that $\forall \varepsilon > 0, \exists \varepsilon' \in (0, \varepsilon)$ s.t. $\chi_{[\delta_f - \varepsilon', \delta_f)} \in f(u)$ thus $f_{RI}^{\mu_r,\delta_r,\mu_f,\delta_f}$ is Zeno, contradiction. Similarly, the supposition $\delta_r \leq \delta_f - \mu_f$ gives a contradiction.

**Corollary 12.4** No Zeno relatively inertial delays $f \subset f_{RI}^{\mu_r,\delta_r,\mu_f,\delta_f}$ exist if $\delta_f > \delta_r - \mu_r, \delta_r > \delta_f - \mu_f$.

**Proof** If $\delta_f > \delta_r - \mu_r, \delta_r > \delta_f - \mu_f$ then $f_{RI}^{\mu_r,\delta_r,\mu_f,\delta_f}$ is not Zeno from Corollary 12.3. Any system $f \subset f_{RI}^{\mu_r,\delta_r,\mu_f,\delta_f}$ is non-Zeno.

Author's address:
str. Zimbrului, Nr. 3, Bl. PB68, Ap. 11, 410430, Oradea
serban_e_vlad@yahoo.com, www.geocities.com/serban_e_vlad